\documentclass[
prl
,byrevtex
,twocolumn
,showkeys]{revtex4}
\usepackage[LY1]{fontenc}
\usepackage{amsmath,amssymb,amsxtra}
\usepackage[dvips]{graphicx}
\usepackage{float}
\newcommand{\bs}[1]{\boldsymbol{#1}}

\begin{document}
\title{Accelerating Universe via Spatial Averaging}
\author{Yasusada Nambu}
\email{nambu@gravity.phys.nagoya-u.ac.jp}
\author{Masayuki Tanimoto}
\email{tanimoto@gravity.phys.nagoya-u.ac.jp}
\affiliation{Department of Physics, Graduate School of Science, Nagoya 
University, Chikusa, Nagoya 464-8602, Japan}
\preprint{DPNU-05-12}
\date{June 14, 2005}
\begin{abstract}
  We present a model of an inhomogeneous universe that leads to
  accelerated expansion after taking spatial averaging. The model
  universe is the Tolman-Bondi solution of the Einstein equation and
  contains both a region with positive spatial curvature and a region
  with negative spatial curvature. We find that after the region with
  positive spatial curvature begins to re-collapse, the deceleration
  parameter of the spatially averaged universe becomes negative and
  the averaged universe starts accelerated expansion. We also discuss the
  generality of the condition for accelerated expansion of the
  spatially averaged universe.
\end{abstract}
\maketitle
\paragraph{Introduction.---}%
The existence of the dark energy component  is considered to be
necessary to explain  present acceleration of the Universe indicated
by recent observational data\cite{BahcallNA:S284:1999,
  PeeblesPJE:RMP75:2003}. 
Although a cosmological constant or a negative pressure fluid are candidates
for the dark energy, we do not know the true character of them and
they are left as black boxes.

More conservative approach to explain the acceleration of the Universe
without introduction of exotic fields is to utilize a backreaction
effect due to inhomogeneities of the
Universe\cite{MukhanovVM:1997,NambuY:2002.1,RasanenS:0311257:2003,KolbEW:0503117:2005,KolbEW:0506534:2005,BuchertT:0507028:2005}. In
this approach, we have to fit a homogeneous and isotropic
Friedmann-Robertson-Walker (FRW) universe to an inhomogeneous universe\cite{EllisGFR:CQG4:1987}.
This is done by taking spatial average of fluctuations of metric and
matter fields. The effect of the inhomogeneities modifies the
evolution of spatially averaged ``background'' FRW universe.

Many researchers have tried to apply this idea to the dark energy
problem, no one has yet succeeded to obtain an workable example that
explains the accelerated expansion of the Universe. This is due to the
non-linearity of the Einstein equation. Usually, the backreaction
effect in the Universe is evaluated by cosmological perturbation
theory. However, after density fluctuation in the Universe grows to be
non-linear and begins to re-collapse, perturbative expansion breaks
down and we cannot obtain reliable results beyond this time based on
perturbative calculation\footnote{The renormalization group method
  improve the secular divergence of the perturbative expansion that
  appears in course of  evolution of density fluctuations and
  circumvents the situation\cite{NambuY:1999}. }. Thus, we expect that
non-perturbative feature of inhomogeneity is crucial to obtain a
workable model that explains the accelerated expansion of the Universe
by the backreaction effect.

In this letter, we consider a model of inhomogeneous universe that is
 an  exact solution of the Einstein equation with dust field. By
 taking spatial average of this model explicitly, we obtained 
 an accelerated expansion  of the spatially averaged universe
 provided that the following two conditions are satisfied:
 \begin{enumerate}
   \item The universe contains both a region with positive spatial
   curvature and a region with negative spatial curvature.
   \item The region with positive spatial curvature is in contracting phase.
 \end{enumerate}
 The degree of the acceleration of the averaged universe depends on
 the scale of the initial inhomogeneity.  We use units in which
 $c=\hbar=8\pi G=1$ throughout the letter.

\paragraph{Averaging of Tolman-Bondi model.---}
We consider a spherically symmetric  solution of the Einstein equation
with dust field. This is the Lema\^{i}tre-Tolman-Bondi
solution\cite{KrasinskiA:CUP:1997} and the  metric is given by
\begin{align}
  &ds^2=-dt^2+\frac{(R_{,r})^2}{1+2E(r)}dr^2+R^2d\Omega_2^2, \label{eq:metric}\\
  & \left(\frac{\dot
      R}{R}\right)^2=\frac{2E(r)}{R^2}+\frac{2M(r)}{R^3}, \label{eq:eq-R}
\end{align}
where $E(r)$ and $M(r)$ are arbitrary functions of $r$. The solution
of Eq.~\eqref{eq:eq-R} can be written parametrically by using a  
variable $\eta=\int dt/R$,
\begin{align}
  &R(\eta,
  r)=\frac{M(r)}{-2E(r)}\left[1-\cos\left(\sqrt{-2E(r)}\,\eta\right)\right],
  \notag\\
  &t(\eta, r)=\frac{M(r)}{-2E(r)}\left[
    \eta-\frac{1}{\sqrt{-2E(r)}}\sin\left(\sqrt{-2E(r)}\,\eta\right)\right].
\end{align}
By introducing the following variables 
\begin{equation}
  a(t,r)=\frac{R(t,r)}{r},\quad k(r)=-\frac{2E(r)}{r^2},\quad
  \rho_0(r)=\frac{6M(r)}{r^3},
\end{equation}
the metric and the evolution equation for the scale factor $a(t,r)$ become
\begin{align}
  &ds^2=-dt^2+a^2\left[\left(1+\frac{a_{,r}r}{a}\right)^2
    \frac{dr^2}{1-k(r)r^2}+r^2d\Omega_2^2\right], \\
  &\left(\frac{\dot
      a}{a}\right)^2=-\frac{k(r)}{a^2}+\frac{\rho_0(r)}{3a^3}. \label{eq:FReq}
\end{align}
Eq.~\eqref{eq:FReq} is same as the Friedmann equation with dust and we can
regard the Tolman-Bondi solution as a model of  inhomogeneous universe
of which local behavior is equivalent to a FRW universe with a
curvature constant $k(r)$. 

\begin{figure}[H]
  \centering
  \includegraphics[width=1.0\linewidth,clip]{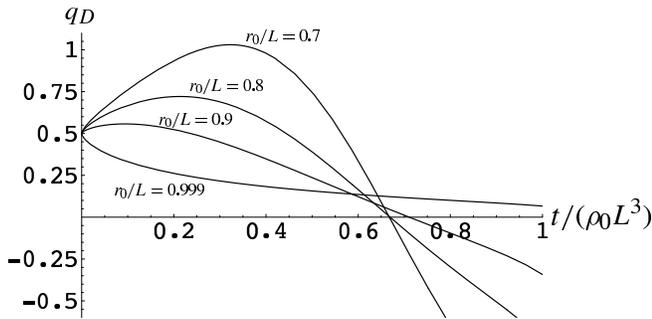}%
  \caption{Evolution of the deceleration parameter $q_D$ of the
    spatially averaged universe for $r_0/L=0.999, 0.9,0.8,0.7$.}
  \label{fig:fig1}
\end{figure}

As a specific case,  we assume the
following spatial distribution of spatial curvature,
\begin{equation}
  k(r)=\frac{1}{L^2}\left[2\theta(r-r_0)-1\right],\quad 0\le r\le
  L,\quad 0\le r_0\le L
\end{equation}
and assume that $\rho_0(r)=\rho_0=$constant. 
In the region 1 : $0\le r<r_0$, the solution is that of a spatially open FRW universe
and in the region 2 : $r_0<r\le L$, the solution is that of a spatially closed
FRW universe. The spatial volume of the comoving region $D: 0\le r\le L$ is 
\begin{align}
  V_D(t)&=\int_0^Ldr\frac{r^2}{\sqrt{1-k(r)r^2}}\left(1+\frac{a_{,r}r}{a}\right)a^3
  \notag \\
  &=L^3\left(c_1a_1^3+c_2a_2^3\right),
\end{align}
where
\begin{align}
  &c_1=\int_0^{r_0/L}\frac{x^2dx}{\sqrt{1+x^2}},\quad
   c_2=\int_{r_0/L}^1\frac{x^2dx}{\sqrt{1-x^2}}
\end{align}
and the scale factor $a_1, a_2$ obey the following Friedmann equations:
\begin{equation}
  \left(\frac{\dot
      a_1}{a_1}\right)^2=+\frac{1}{L^2a_1^2}+\frac{\rho_0}{3a_1^3},\quad
 \left(\frac{\dot a_2}{a_2}\right)^2=-\frac{1}{L^2a_2^2}+\frac{\rho_0}{3a_2^3}.
\end{equation}
We define a spatially averaged scale factor of the region $D$ by its
physical volume
\begin{equation}
  a_D\equiv\left(\frac{V_D(t)}{V_D(t_*)}\right)^{1/3}, \label{eq:ad}
\end{equation}
where we have normalized $a_D=1$ at a fixed time $t=t_*$. We can evaluate the deceleration
parameter of the spatially averaged scale factor
\begin{equation}
  q_D=-\frac{\ddot a_Da_D}{\dot a_D^2}.
\end{equation}
Fig.~\ref{fig:fig1} shows evolution of $q_D$ for different values of
$r_0/L$.  About $t\sim 0$, the averaged universe behaves as the
spatially flat FRW model with dust. The averaged universe enters the
accelerating phase at a certain time after the closed FRW region
begins to re-collapse at $t=\pi\rho_0L^3/6$%
\footnote{The reader might wonder why the existence of the contracting
region could help accelerate the (averaged ) universe. This is however
the case, as seen from the following identity
$$
X^2\ddot X=2b_1b_2\frac{(\dot b_1b_2-b_1\dot b_2)^2}{b_1^3+b_2^3}+
 b_1^2\ddot b_1+b_2^2\ddot b_2,
$$
where $X=(b_1^3+b_2^3)^{1/3}$, and $b_1$ and $b_2$ are arbitrary
functions of time. In this identity if we regard
$b_i=L(c_i/V_*)^{1/3}a_i~ (i=1,2)$, $X$ corresponds to the averaged
scale factor $a_D$, and one can easily see that even if each
acceleration $\ddot b_i$ is negative, the acceleration $\ddot X$ for
the averaged one will become positive if $\dot b_1$ and $\dot b_2$
have the opposite signs (meaning one is expanding and the other is
contracting) and the first positive term in the identity becomes
large. This is exactly what is happening in the model.}.
 As the ratio $r_0/L$ becomes smaller, the
averaged universe enters the accelerating phase earlier.  For too small
ratio $r_0/L<0.4$, the averaged universe re-collapse and we do not
have the accelerated expansion.

\paragraph{The condition for accelerated expansion.---}
To obtain the condition for accelerated expansion of the spatially
averaged universe, we consider a model of inhomogeneous universe of
which local scale factor $a(t,\bs{x})$ obeys the following Einstein
equation :
\begin{align}
  &\left(\frac{\dot
      a}{a}\right)^2=-\frac{k(\bs{x})}{a^2}+\frac{\rho}{3}, \\
  &\frac{\ddot a}{a}=-\frac{\rho}{6},\quad \dot\rho+3\left(\frac{\dot a}{a}\right)\rho=0.
\end{align}
This model includes the Tolman-Bondi solution presented previously. We assume
that the volume of the spatial region $D$ is given by
\begin{equation}
  V_D=\int_Dd^3x\sqrt{h}\,a^3(t,\bs{x})
\end{equation}
where $\sqrt{h}$ is a time independent function of spatial coordinate
$\bs{x}$ and its explicit form is not necessary for the following
analysis.  The spatially averaged scale factor is defined by
Eq.~\eqref{eq:ad} and the spatial average of a quantity $f(t,\bs{x})$ is
defined by
\begin{equation}
  \langle f\rangle\equiv \frac{1}{V_D}\int_Dd^3x \sqrt{h}\, a^3 f.
\end{equation}
The important property of the operation of the spatial averaging is that it does not
commute with the time derivative\cite{BuchertT:GRG32:2000}:
\begin{equation}
  \langle f\rangle\spdot=\langle\dot f\rangle+3\left[\langle
    fH\rangle-\langle H\rangle\langle f\rangle\right].
\end{equation}
The average of the local Hubble parameter $H=\dot a/a$ is
\begin{equation}
  H_D\equiv\langle H\rangle=\frac{\dot a_D}{a_D}.
\end{equation}
The Einstein equation for the averaged scale factor becomes
\begin{align}
  &\left(\frac{\dot
      a_D}{a_D}\right)^2=\frac{\rho_D}{3}-\mathcal{R}-\mathcal{Q}, \label{eq:ein-1}\\
  &\frac{\ddot a_D}{a_D}=-\frac{\rho_D}{6}+2\mathcal{Q},\quad
  \dot\rho_D+3H_D\rho_D=0,
\label{eq:ein-2}
\end{align}
where we have defined
\begin{equation}
  \mathcal{R}=\left\langle\frac{k(\bs{x})}{a^2}\right\rangle,\quad
  \mathcal{Q}=\langle H^2\rangle-H_D^2=\frac{\rho_D}{3}-\mathcal{R}-H_D^2.
\end{equation}
The integrability condition \cite{BuchertT:GRG32:2000} of Eq.~\eqref{eq:ein-1} and
Eq.~\eqref{eq:ein-2} is
\begin{equation}
  a_D^4(a_D^2\mathcal{R})\spdot+(a_D^6\mathcal{Q})\spdot=0
\end{equation}
and this equation does not yield independent evolution equation. Thus,
we can integrate Eqs.~\eqref{eq:ein-1} and \eqref{eq:ein-2} by
assuming that the scale factor dependence of the averaged spatial
curvature $\mathcal{R}=\mathcal{R}(a_D)$. After all, the Einstein
equation for the averaged scale factor becomes
\begin{align}
  &\left(\frac{\dot
      a_D}{a_D}\right)^2=\frac{\rho_D}{3}-\frac{4}{a_D^6}
  \int_1^{a_D}dxx^5\mathcal{R}(x), \label{eq:ein-ave1}\\
  &\frac{\ddot a_D}{a_D}=-\frac{\rho_D}{6}-2\mathcal{R}(a_D)
  +\frac{8}{a_D^6}\int_1^{a_D}dxx^5\mathcal{R}(x).
  \label{eq:ein-ave2}
\end{align}
The averaged dust density is $\rho_D=\rho_*/a_D^3$
where $\rho_*=\rho_D(a_D=1)$. 
We can solve these equations if we obtain the scale factor dependence
of the averaged spatial curvature $\mathcal{R}$.

For the Tolman-Bondi model introduced previously, the averaged
dust density is given by
\begin{equation}
  \rho_D=\left(\frac{\rho_0L^3}{V_*}\right)\frac{c_1+c_2}{a_D^3},\quad V_*=V_D(a_D=1)
\end{equation}
and the averaged spatial curvature is
\begin{equation}
  \mathcal{R}=\frac{L}{V_*}\left(\frac{-c_1a_1+c_2a_2}{a_D^3}\right).
\end{equation}
We evaluate $\mathcal{R}$ about the time at which the closed FRW
region (region 2) begins to re-collapse. The scale factor of the region
2 stays nearly constant $a_2\approx \rho_0L^2/3\ll a_1$ around the
maximal expansion and the averaged spatial curvature
is approximated to be\footnote{The present approximation serves a good
estimate of the coefficients $K$ and $C$ for smaller $r_0/L$. When
this ratio is closer to unity, we can determine the coefficients
exactly by means of a comparison of Taylor expansion.}
\begin{equation}
  \mathcal{R}\approx \frac{K}{a_D^2}+\frac{C}{a_D^3}, \label{eq:R}
\end{equation}
where
\begin{equation}
  K=-\left(\frac{c_1}{V_*}\right)^{2/3},\quad
  C=\frac{\rho_0L^3c_2}{3V_*}.
\end{equation}
By substituting \eqref{eq:R} to Eqs.~\eqref{eq:ein-ave1} and
\eqref{eq:ein-ave2}, we obtain
\begin{align}
  \left(\frac{\dot
      a_D}{a_D}\right)^2&=\left(\frac{\rho_0}{3}-\frac{4C}{3}\right)\frac{1}{a_D^3} 
  -\frac{K}{a_D^2}+\left(K+\frac{4C}{3}\right)\frac{1}{a_D^6}\notag\\&\equiv
  \frac{\rho_{\text{eff}}}{3}, \\
  \frac{\ddot
    a_D}{a_D}&=\left(-\frac{\rho_0}{6}+\frac{2C}{3}\right)\frac{1}{a_D^3}
   -2\left(K+\frac{4C}{3}\right)\frac{1}{a_D^6}\notag \\&\equiv
   -\frac{1}{6}(\rho_{\text{eff}}+3p_{\text{eff}}),
\end{align}
and for $a_D\gg 1$,
\begin{equation}
  \rho_{\text{eff}}\approx \frac{\rho_*-4C}{a_D^3}-\frac{3K}{a_D^2},\quad 
  p_{\text{eff}}\approx \frac{K}{a_D^2}.
\end{equation}
Asymptotically, the equation of state of the averaged universe is
\begin{equation}
  w_{\text{eff}}=\frac{p_{\text{eff}}}{\rho_{\text{eff}}}\approx
  -\frac{1}{3}-\left(\frac{4C-\rho_*}{-3K}\right)\frac{1}{a_D}.
\end{equation}

The condition for the accelerated expansion of the universe is
$\rho_{\text{eff}}>0$ and $\rho_{\text{eff}}+3p_{\text{eff}}<0$ and
this yields the following relation between constants contained in our model: 
\begin{equation}
  K<0,\quad 0<4C-\rho_*<3(-K). \label{eq:condi}
\end{equation}
For the Tolman-Bondi model, the first condition $K<0$ is automatically
satisfied. This condition means that the average of spatial curvature
in the inhomogeneous universe must be negative asymptotically in
time. That is, the comoving region $D$ must contains a spatially open
region.  The condition $0<4C-\rho_*$ corresponds to violation of the
strong energy condition of the averaged universe, and it yields
\begin{equation}
  c_1<c_2/3 \label{eq:cond-1}.
\end{equation}
The condition $4C-\rho_*<3(-K)$ means the averaged universe does
re-collapse before it enters the accelerating phase and yields
\begin{equation}
  \label{eq:cond-2}
  (c_2/3-c_1)c_1^{-2/3}<\frac{3}{\rho_0L^2}\left(\frac{V_*}{L^3}\right)^{1/3}.
\end{equation}
Thus, the conditions \eqref{eq:cond-1} and \eqref{eq:cond-2} constrain the
size of spatially open region. For $\rho_0 L^2=1, V_*=L^3$, we have
\begin{equation}
  0.4<r_0/L<0.9,
\end{equation}
and this range of the parameter is consistent with the behavior of the
Tolman-Bondi model. 
If the spatially open region is too large, the averaged universe can
not enter accelerating phase before the spatially closed region
re-collapse. On the other hand, if the spatially open region is too
small, the averaged universe will re-collapse before it enters the
accelerating phase.

\paragraph{Conclusion.---}

In this letter, we have presented an example of inhomogeneous universe that
shows the accelerated expansion after taking spatial averaging. The necessary
condition to realize accelerated expansion is that the inhomogeneous
universe contains both a spatially open region (with negative spatial
curvature)  and a spatially closed region (with positive spatial
curvature). Although if the whole universe starts its evolution with
positive expansion, the region with positive spatial curvature  will
re-collapse after all. The scale factor dependence of the averaged
curvature $\mathcal{R}$ in Eq.~\eqref{eq:R}, that is crucial to
derive the accelerated expansion, is the result of co-existence of the
expanding spatial region and the contracting spatial region. This is
the non-perturbative feature of inhomogeneity and it is not possible
to derive the same result based on the ordinary perturbative
calculation. Even though we have checked this form of the averaged
spatial curvature only for a  spherically symmetric case, we expect
that this form of the averaged spatially  curvature will hold for more
general case without symmetry and the acceleration of the averaged
universe is a general feature of inhomogeneous universe with non-linear
fluctuations.

As  an application of our model to the Universe, it is possible to fix the scale
of inhomogeneity $r_0, L$ by using the present value of the
cosmological parameters. We will report on this subject in a
forthcoming paper.

\begin{acknowledgments}
This work was supported in part by a Grant-In-Aid for Scientific
Research of the Ministry of Education, Science, Sports, and Culture of
Japan (11640270).
\end{acknowledgments}


\begin{thebibliography}{12}
\expandafter\ifx\csname natexlab\endcsname\relax\def\natexlab#1{#1}\fi
\expandafter\ifx\csname bibnamefont\endcsname\relax
  \def\bibnamefont#1{#1}\fi
\expandafter\ifx\csname bibfnamefont\endcsname\relax
  \def\bibfnamefont#1{#1}\fi
\expandafter\ifx\csname citenamefont\endcsname\relax
  \def\citenamefont#1{#1}\fi
\expandafter\ifx\csname url\endcsname\relax
  \def\url#1{\texttt{#1}}\fi
\expandafter\ifx\csname urlprefix\endcsname\relax\def\urlprefix{URL }\fi
\providecommand{\bibinfo}[2]{#2}
\providecommand{\eprint}[2][]{\url{#2}}

\bibitem[{\citenamefont{Bahcall et~al.}(1999)\citenamefont{Bahcall, Ostriker,
  Perlmutter, and Steinhardt}}]{BahcallNA:S284:1999}
\bibinfo{author}{\bibfnamefont{N.~A.} \bibnamefont{Bahcall}},
  \bibinfo{author}{\bibfnamefont{J.~P.} \bibnamefont{Ostriker}},
  \bibinfo{author}{\bibfnamefont{S.}~\bibnamefont{Perlmutter}},
  \bibnamefont{and} \bibinfo{author}{\bibfnamefont{P.~J.}
  \bibnamefont{Steinhardt}}, \bibinfo{journal}{Science}
  \textbf{\bibinfo{volume}{284}}, \bibinfo{pages}{1481} (\bibinfo{year}{1999}).

\bibitem[{\citenamefont{Peebles and Ratra}(2003)}]{PeeblesPJE:RMP75:2003}
\bibinfo{author}{\bibfnamefont{P.~J.~E.} \bibnamefont{Peebles}}
  \bibnamefont{and} \bibinfo{author}{\bibfnamefont{B.}~\bibnamefont{Ratra}},
  \bibinfo{journal}{Rev. Mod. Phys.} \textbf{\bibinfo{volume}{75}},
  \bibinfo{pages}{559} (\bibinfo{year}{2003}).

\bibitem[{\citenamefont{Buchert}(2000)}]{BuchertT:GRG32:2000}
\bibinfo{author}{\bibfnamefont{T.}~\bibnamefont{Buchert}},
  \bibinfo{journal}{Gen. Rel. Grav.} \textbf{\bibinfo{volume}{32}}, \bibinfo{pages}{105}
  (\bibinfo{year}{2000}).

\bibitem[{\citenamefont{Mukhanov et~al.}(1997)\citenamefont{Mukhanov, Abramo,
  and Brandenberger}}]{MukhanovVM:1997}
\bibinfo{author}{\bibfnamefont{V.~F.} \bibnamefont{Mukhanov}},
  \bibinfo{author}{\bibfnamefont{L.~R.} \bibnamefont{Abramo}},
  \bibnamefont{and} \bibinfo{author}{\bibfnamefont{R.~H.}
  \bibnamefont{Brandenberger}}, \bibinfo{journal}{Phys. Rev. Lett.}
  \textbf{\bibinfo{volume}{78}}, \bibinfo{pages}{1624} (\bibinfo{year}{1997}).

\bibitem[{\citenamefont{Nambu}(2002)}]{NambuY:2002.1}
\bibinfo{author}{\bibfnamefont{Y.}~\bibnamefont{Nambu}},
  \bibinfo{journal}{Phys. Rev. D} \textbf{\bibinfo{volume}{65}},
  \bibinfo{pages}{104013} (\bibinfo{year}{2002}).

\bibitem[{\citenamefont{R{\"a}s{\"a}nen}(2003)}]{RasanenS:0311257:2003}
\bibinfo{author}{\bibfnamefont{S.}~\bibnamefont{R{\"a}s{\"a}nen}},
  \bibinfo{journal}{astro-ph/0311257}  (\bibinfo{year}{2003}).

\bibitem[{\citenamefont{Kolb et~al.}(2005{\natexlab{a}})\citenamefont{Kolb,
  Matarrese, Notari, and Riotto}}]{KolbEW:0503117:2005}
\bibinfo{author}{\bibfnamefont{E.~W.} \bibnamefont{Kolb}},
  \bibinfo{author}{\bibfnamefont{S.}~\bibnamefont{Matarrese}},
  \bibinfo{author}{\bibfnamefont{A.}~\bibnamefont{Notari}}, \bibnamefont{and}
  \bibinfo{author}{\bibfnamefont{A.}~\bibnamefont{Riotto}},
  \bibinfo{journal}{hep-th/0503117}  (\bibinfo{year}{2005}{\natexlab{a}}).

\bibitem[{\citenamefont{Kolb et~al.}(2005{\natexlab{b}})\citenamefont{Kolb,
  Matarrese, and Riotto}}]{KolbEW:0506534:2005}
\bibinfo{author}{\bibfnamefont{E.~W.} \bibnamefont{Kolb}},
  \bibinfo{author}{\bibfnamefont{S.}~\bibnamefont{Matarrese}},
  \bibnamefont{and} \bibinfo{author}{\bibfnamefont{A.}~\bibnamefont{Riotto}},
  \bibinfo{journal}{astro-ph/0506534}  (\bibinfo{year}{2005}{\natexlab{b}}).

\bibitem[{\citenamefont{Buchert}(2005)}]{BuchertT:0507028:2005}
\bibinfo{author}{\bibfnamefont{T.}~\bibnamefont{Buchert}},
  \bibinfo{journal}{gr-ac/0507028}  (\bibinfo{year}{2005}).


\bibitem[{\citenamefont{Ellis and Stoeger}(1987)}]{EllisGFR:CQG4:1987}
\bibinfo{author}{\bibfnamefont{G.~F.~R.} \bibnamefont{Ellis}} \bibnamefont{and}
  \bibinfo{author}{\bibfnamefont{W.}~\bibnamefont{Stoeger}},
  \bibinfo{journal}{Class. Quant. Grav.} \textbf{\bibinfo{volume}{4}},
  \bibinfo{pages}{1687} (\bibinfo{year}{1987}).



\bibitem[{\citenamefont{Nambu and Yamaguchi}(1999)}]{NambuY:1999}
\bibinfo{author}{\bibfnamefont{Y.}~\bibnamefont{Nambu}} \bibnamefont{and}
  \bibinfo{author}{\bibfnamefont{Y.~Y.} \bibnamefont{Yamaguchi}},
  \bibinfo{journal}{Phys. Rev. D} \textbf{\bibinfo{volume}{60}},
  \bibinfo{pages}{104011} (\bibinfo{year}{1999}).

\bibitem[{\citenamefont{Krasi{\'n}ski}(1997)}]{KrasinskiA:CUP:1997}
\bibinfo{author}{\bibfnamefont{A.}~\bibnamefont{Krasi{\'n}ski}},
  \emph{\bibinfo{title}{Inhomogeneous cosmological models}}
  (\bibinfo{publisher}{Cambridge University Press}, \bibinfo{year}{1997}), ISBN
  \bibinfo{isbn}{0-521-48180-5}.

\end{thebibliography}

\end{document}